# Maxwell's demon for calcium binding to calmodulin?

S. L. Mironov

*Institute of Neuro- and Sensory Physiology, Georg-August-University, Göttingen 37073, Germany*

Tel.: +49 (551) 39 54 72; E-mail: smirono@gwdg.de.

In the recent paper (2011) Faas and co-workers claimed that calmodulin could directly detect $Ca^{2+}$ signals by acting as extremely fast calcium buffer. However the results and their interpretation raise serious doubts about this conclusion.

The major concerns are:

- The on-rate constant for calcium binding by calmodulin (CaM) has not been directly measured and was instead found via Markovian fitting of the complex scheme that includes many steps. Having obtained via fitting an extremely fast time-constant for calcium-CaM binding, the authors intended to substantiate this by comparing the estimate with diffusion limit. As shown by Einstein and Smoluchowsky the latter determines the maximally possible rate constant of reaction between two particles from the frequency of their collisions and limits the on-rate constant from above. The authors argue that the diffusion limit for Ca-CaM binding may correspond to the rate constant $5 \cdot 10^{10}$ $M^{-1}s^{-1}$ that corroborates the results of fitting. The estimate is however much higher than ever reported for calcium binding to other molecules as in the case of ATP considered below.

- Furthermore, the time-resolution of experiments performed is not fast enough to resolve such time-constants. The first measurement point in Fig. 1d in (Faas et al. 2011) is ~0.2 ms and the authors show a decay of calcium signal with a half-time ~0.3 ms. Assuming that the on-rate constant is indeed $3 \cdot 10^{10}$ $M^{-1}s^{-1}$ and multiplying it by calmodulin concentration (14 µM) in the experiment, one gets the first order rate-constant $4 \cdot 10^{5}$ $s^{-1}$ that corresponds to the time constant ~2.5 µs i. e. 100-fold smaller than the measurements show.

- Using Einstein-Smoluchowsky formula to obtain the diffusion limit, Faas et al. set the radius of calmodulin to 2.5 nm. It is a bulky molecule containing four calcium binding pockets whose capture radius is much smaller, in the range of ionic radius for calcium (0.1 nm). The proper estimate of diffusion limit for Ca-CaM binding should be accordingly smaller, likely close to the measured half-decay time.

- The authors considered also calcium binding by ATP, but set the on-rate constant to $2 \cdot 10^8 \, M^{-1} s^{-1}$ in this case. Both ATP and CaM have 4 negative charges that participate in calcium binding. Therefore their capture radii for calcium and corresponding on-rate constants should be about the same, below theoretical diffusion limit.

In conclusion,

- The experimental approach used by Faas et al. (2011) does not have sufficient time-resolution (in µs range) to measure the on-rate constants bigger than $10^8 \, M^{-1} s^{-1}$;
- The diffusion limit for Ca-CaM binding was not correctly estimated.